\documentclass{raa_rb}            

\usepackage{graphicx,times}             
\usepackage{natbib}
\usepackage{amssymb,amsmath}
\bibpunct{(}{)}{;}{a}{}{,}

\usepackage[pagebackref=true]{hyperref}

\begin{document}

  \title{About 300 days optical quasi-periodic oscillations in the long-term light curves of the blazar PKS~2155-304
}

   \volnopage{Vol.0 (20xx) No.0, 000--000}      
   \setcounter{page}{1}          

   \author{Qi Zheng 
      \inst{1}
   \and XueGuang Zhang
      \inst{2}
   \and QiRong Yuan
      \inst{3}
   }

   \institute{School of Physics and technology, Nanjing Normal University, No. 1, 
   	Wenyuan Road, Nanjing, 210023, P. R. China; {\it 211002011@njnu.edu.cn}\\
        \and
             School of Physics and technology, Nanjing Normal University, No. 1, 
             Wenyuan Road, Nanjing, 210023, P. R. China; {\it xgzhang@njnu.edu.cn}\\
        \and
             School of Physics and technology, Nanjing Normal University, No. 1, 
             Wenyuan Road, Nanjing, 210023, P. R. China; {\it yuanqr@njnu.edu.cn}\\
\vs\no
   {\small Received 20xx month day; accepted 20xx month day}}

\abstract{Based on the long-term light curves collected from the Catalina Sky Survey (CSS) (from 
	2005 to 2013) and the All-Sky Automated Survey for Supernovae (ASAS-SN) (from 2014 to 2018), 
	optical quasi-periodic oscillations (QPOs) about 300 days can be well determined in the well-known 
	blazar PKS~2155-304 through four different methods: the generalized Lomb-Scargle periodogram 
	(GLSP) method, the weighted wavelet Z-transform (WWZ) technique, the epoch-folded method and redfit method. 
    The GLSP determined significance level for the periodicity is higher 
	than 99.9999\% based on a false alarm probability.
	The redfit provided confidence level for the periodicity is higher than 99\% 
	in ASAS-SN light curve, after considering the effects of red noise.
	Based on continuous autoregressive (CAR) process created artificial light curves, the probability of detecting fake QPOs is lower than 0.8\%.
	The determined optical periodicity of 300 days from CSS and  ASAS-SN light curves is well consistent 
	with the reported optical periodicity in the literature.
	Moreover, three possible models are discussed to explain the optical QPOs in PKS 2155-304: 
	the relativistic Frame-dragging effect, the binary black hole (BBH) model and the jet precession model.
\keywords{galaxies:active - galaxies:BL Lacertae objects - individual:PKS~2155-304:quasi-periodic oscillation}
}

   \authorrunning{Q. Zheng,X. G. Zhang \&Q. R. Yuan}            
   \titlerunning{Optical QPOs in PKS~2155-304}  

   \maketitle

%
%
\section{Introduction}

PKS 2155–304 ({$z = 0.116$}) \citep{Ah09} is one of the best known blazars and one of the 
brightest objects from the UV to TeV energies in the southern sky \citep{Ca92,Fo07} . With the Parkes 
survey, PKS~2155-304 has been observed in radio band \citep{Sh74}. Using HEAo-1, \citet{Sc79} 
completed its first X-ray observations. Due to its strong and variable X-ray emissions, it was 
classified as a BL Lac object \citep{ZG21}. PKS 2155–304 was first identified as a TeV blazar by 
the detection of VHE gamma rays by the Durham MK 6 telescopes \citep{Ch99}, and then was confirmed 
by the H.E.S.S. \citep{Ah05}. PKS~2155-304 has been observed on diverse timescales over a wide 
range of frequencies from radio to VHE {\ensuremath{\gamma}}-rays, and shown rapid and strong 
variability \citep{Mi83,Fa00}. Generally, blazar variability timescale ({$t_{\rm var}$}) is divided 
into three classes \citep{Gu04,Ag19}: microvariability (intra-night variability or intra-day 
variability) (IDV; {$t_{\rm var}$}$\sim$ less than a day) \citep{Wa95}, short-term variability 
(STV; {$t_{\rm var}$}$\sim$ from days to few months) and long-term variability (LTV; 
{$t_{\rm var}$}$\sim$ from months to several years) \citep{Pa20}. Through AGN variability, we can 
obtain information with respect to their nature. It is important for quasar modelling \citep{Fa98}. 
Optical variability of PKS~2155-304 has been studied for many years.  PKS~2155-304 has obvious IDV \citep{Pa97,To01,Do06}, STV \citep{Ca92,Pe97} and LTV \citep{Ka11} in optical band, but it also can stay in a completely stable state for one week \citep{He97}. In addition, the optical variability is related to X-ray \citep{Do04} and NIR band \citep{Li18}.
The main focus of this paper is long-term variability in optical V-band in PKS~2155-304.

The quasi-periodic oscillations (QPOs) in optical band have been found in PKS~2155-304. 
In V-band, with data from 1970 onwards, \citet{Fa00} reported QPOs of T $\sim$ 4.2 yr and 
T $\sim$ 7.0 yr, and also put forward a possibility of a periodicity of less than 4 yr, but they 
could not confirm this periodicity because of the lack of data. In an available historical archival 
data-set (data from twenty-five different astronomical groups) in R-band collected for 35 yr 
(covering 1979 - 2013), evidence of QPOs with a periodicity of 317 days was reported in \citet{Zh14}. \citet{Ri10} discussed that the long-term QPOs 
in PKS~2155-304 might indicate a binary black hole (BBH) model, leading to a signal of periodicity 
when the secondary BH crosses the disk of the primary BH, similar as the QPOs in OJ 287 \citep{Va08, 
	Ha13, Br20}. In addition, with the VRIJHK photometry, \citet{Sa14} reported QPOs of T $\sim$ 315 
days by using data from the Rapid Eye Mounting Telescope during 2005–2012, which is well consistent 
with the report by \citet{Zh14}. What's more, the overlap of time interval between two works makes 
the periodicity more robust. A few years later, using the extensive HIPPO data taken during 2009 
July 25 to 27, \citet{Pe16} claimed QPOs of T $\sim$ 13 min and T $\sim$ 30 min, which is the first 
evidence of QPOs in the polarization of AGN, and discussed that such fast variations in the optical 
polarization might generate from an emission region, which is comparable in size to radius of 
gravity of the central engine, in Doppler boosted jet. \citet{Sa18} derived data from the Rapid 
Eye Mounting Telescope photometry, SMARTS, the Tuorla Blazar Monitoring Program, and combined with 
data from the Steward Observatory Fermi Blazar Observational Program, ROTSEIII, the All Sky Automated 
Survey robotic telescopes and archival data collected by \citet{Ka11}, and found a periodicity of 
315 $\pm$ 25 days in R-band, which can be caused by relativistic jet instabilities \citep{Pa20, Ka21} 
or chance fluctuation \citep{Sm18, Ho18}. \citet{Ch19} reported that a 700 day-long periodicity is 
found in optical band, as well as in high energy (100 MeV $<$ E $<$ 300 GeV) with the combination of 
SMARTS, RXTE, Swift/XRT, XMM–Newton, Fermi and H.E.S.S. data, which can be explained by a time-dependent 
synchrotron-self Compton (SSC) \citep{Ab10,Ar21,Jo21} model, except for optical band. In this paper, 
further evidence of optical QPOs will be given in PKS~2155-304 through long-term variability from the 
Catalina Sky Survey (CSS) Data Release 2 \citep{Dr09} and All-Sky Automated Survey for Supernovae 
(ASAS-SN)\citep{Sh14,Ko17}.

In other bands, signals of QPOs have also been detected and reported in PKS~2155-304.
\citet{Ur93} reported a $\sim$ 0.7 day QPOs in the ultraviolet from data obtained by the IUE 
satellite throughout 1991 November on a daily basis, and suggested that the flares may cause from 
disturbances propagating along magnetic field in a jet. Unfortunately, with more data achieved 
during the whole month of 1991 November, \citet{Ed95} could not recover the periodicity above with 
more rigorous analysis. Based on the XMM-Newton EPIC/pn detector observation of 24 data sets, 
\citet{La09} got a periodicity of T $\sim$ 4.6 h on May 1, 2006 in the 0.3 - 10 keV. \citet{Ga10} 
claimed that another light curve, rather than the upper one, from XMM-Newton/EPIC displays a weak 
and broad QPOs with a periodicity of 5.5 $\pm$ 1.3 ks. \citet{Sa14} considered the Fermi light curve 
(from 2008 August 6 to 2014 June 9), and pointed out a periodicity peak at T $\sim$ 630 - 640 days, 
which is twice as many as optical and NIR period. \citet{Zh17} found a 1.74 $\pm$ 0.13 year-long  
{\ensuremath{\gamma}}-ray QPOs in Fermi LAT Pass 8 data with the data from August 2008 to October 2016, 
probably associated with relativistic jet instability or the process feeding the jet. What's more, 
a jets-in-jet model was purposed as a plausible reason to explain the TeV flares. There are blobs 
that move relativistically in the jet, which lead to fast-evolving flares \citep{Gi09}. Rapid TeV 
variability can be well explained using a standard SSC approach while taking into account the particle 
evolution and the external light crossing time effects \citep{Ka08}. \citet{Pr17} provided evidence 
of a 644 day-long periodicity in {\ensuremath{\gamma}}-ray band from Fermi-LAT (3FGL). The result 
corresponds to 1.7 yr proposed by \citet{Pe20} with same data. \citet{Bh20} reported a periodicity 
of $\sim$ 610 days from 3FGL. \citet{Ta20} analysed data from the LAT 8-year Source Catalog, 
spanning from August 4 to 2019 April 19, and found a periodicity of 612 $\pm$ 42 days in 
{\ensuremath{\gamma}}-ray band, confirmed by \citet{Zy21}. According to the long timescale, 
\citet{Zy21} considered that the variability originates in accretion disk.

This paper is organized as follows. In Section 2, the acquisition of the magnitude 
measurements, methods to determine the optical QPOs are presented. Our discussions on the optical 
QPOs with a periodicity about 300 days are given in Section 3. Final summaries and conclusions 
are shown in Section 4. Throughout this paper, we have adopted the cosmological parameters of 
$H_{0}=70{\rm km\cdot s}^{-1}{\rm Mpc}^{-1}$, $\Omega_{\Lambda}=0.7$ and $\Omega_{\rm m}=0.3$.

\section{Main Results}

\begin{figure*}
	\centering\includegraphics[width = 7cm,height=3.5cm]{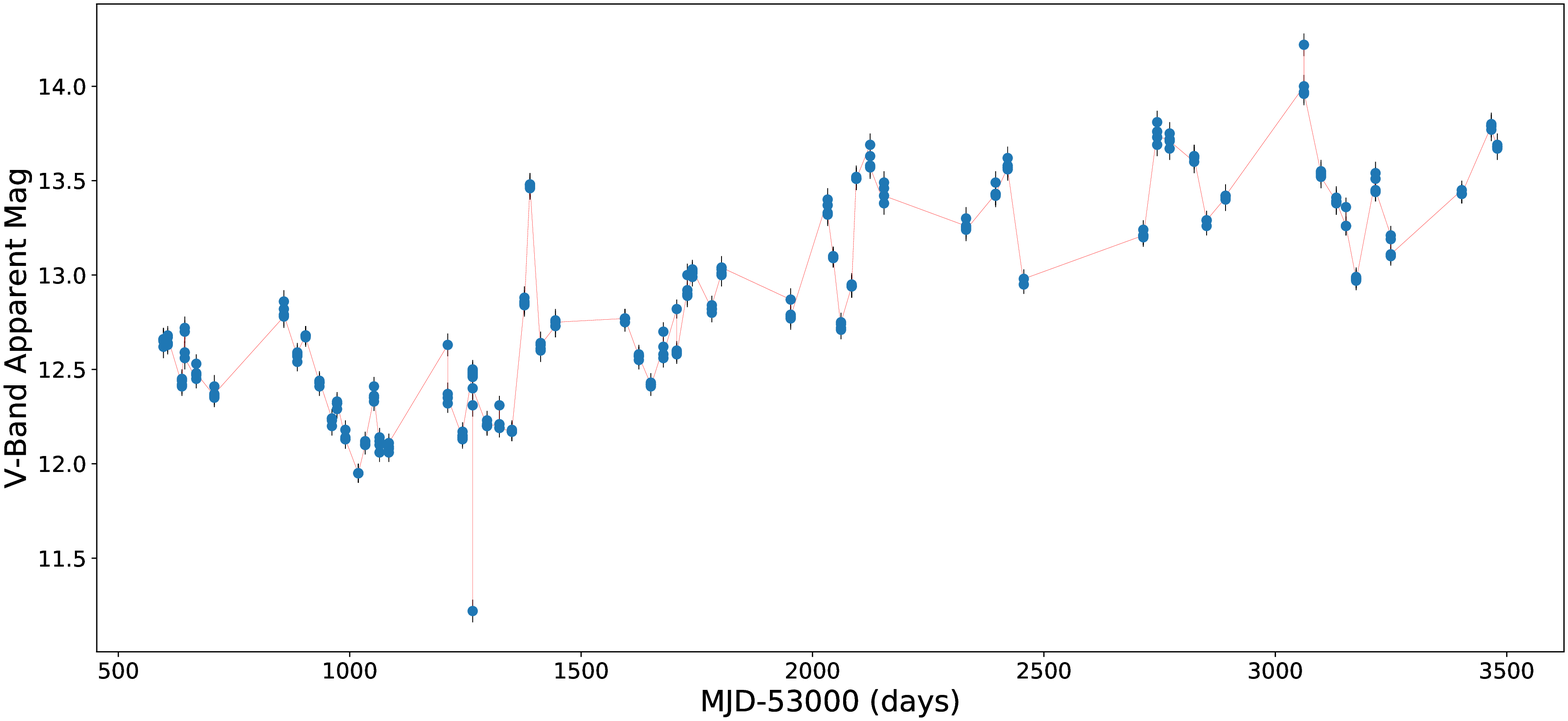}
	\centering\includegraphics[width = 7cm,height=3.5cm]{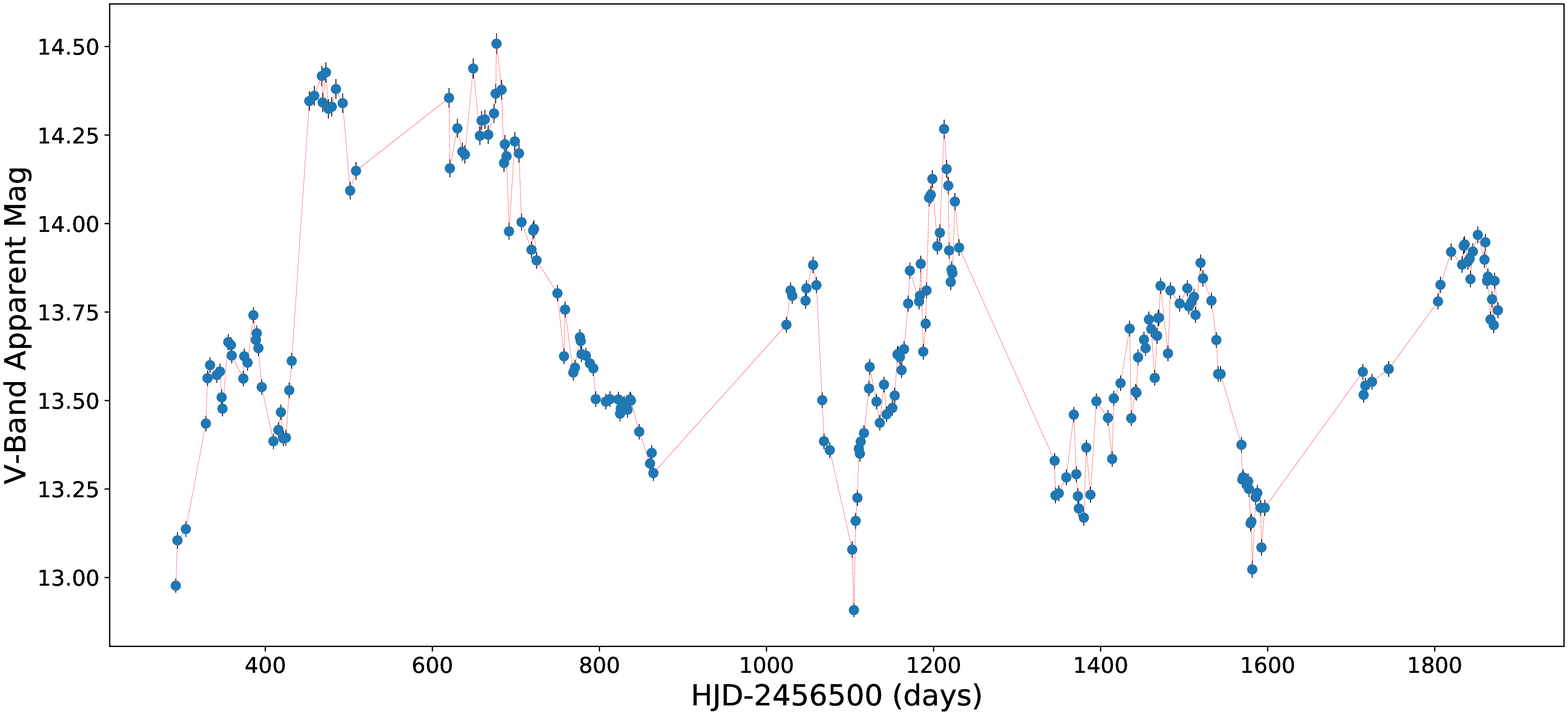}
	\caption{Light curves of PKS~2155-304\ in V-band from CSS (left panel) and in V-band from ASAS-SN 
		(right panel).} 
	\label{fig1}
\end{figure*}

We collected optical-band photometric data of PKS~2155-304 from the Catalina Sky Survey 
(CSS) \citep{Dr09} and from the All-Sky Automated Survey for Supernovae (ASAS-SN) \citep{Sh14,Ko17}, 
with CSS light curve from August 15, 2005 to July 6, 2013 (MJD from 53597.542 to 56479.608), with 
ASAS-SN light curve from May 15, 2014 to September 14, 2018 (HJD from 2456792.777 to 2458375.657), 
shown in Figure \ref{fig1}. 
CSS utilizes three telescopes, 1.5-meter telescope with the field of view $5.0 deg^2$, 
	1.0-meter telescope with the field of view $0.3 deg^2$,
	0.7-meter telescope with the field of view $19.4 deg^2$,
	owned and managed by Steward Observatory of the University of Arizona.
The CSS is a part of the Catalina Surveys. The CSS project is mainly used for searching rapidly moving 
Near Earth Objects (NEOs) and makes efforts to catalogue at least 90 percent of the estimated population of NEOs larger than 140 meters, some of which may pose an impact threat to Earth.
ASAS-SN consists of 24 individual 14-cm telescopes, distributed around the globe with six units located at the Hawaii station of the Las Cumbres Observatory, South Africa, Texas, China and two at Chile.
ASAS-SN currently works in the optical wavelength range and is survey to monitor
daily the entire night sky. 
The V-band used in ASAS-SN is Johnson V-band filter. In addition, ASAS-SN is now focusing on fainter objects in the g-band and the rate of data collection has been improved to 20 hours, which will make ASAS-SN discover more variable objects in greater detail than before.  
Due to unknown magnitude difference between ASAS-SN V-band and g-band 
light curves in PKS~2155-304, and due to quite short time duration of ASAS-SN g-band light curve, 
the ASAS-SN g-band light curve is not taken into account in this paper.

PKS~2155-304 has been observed for many years and shown optical QPOs. In order to test 
the optical QPOs in the CSS and ASAS-SN light curves, the following commonly accepted methods 
are applied: the generalized Lomb-Scargle periodogram (GLSP) method \citep{Br01,Va18}, the weighted 
wavelet Z-transform (WWZ) technique \citep{An13}, the epoch-folded method 
and redfit method \citep{Sc02}.

\begin{figure*}
	\centering\includegraphics[width = 7cm,height=5cm]{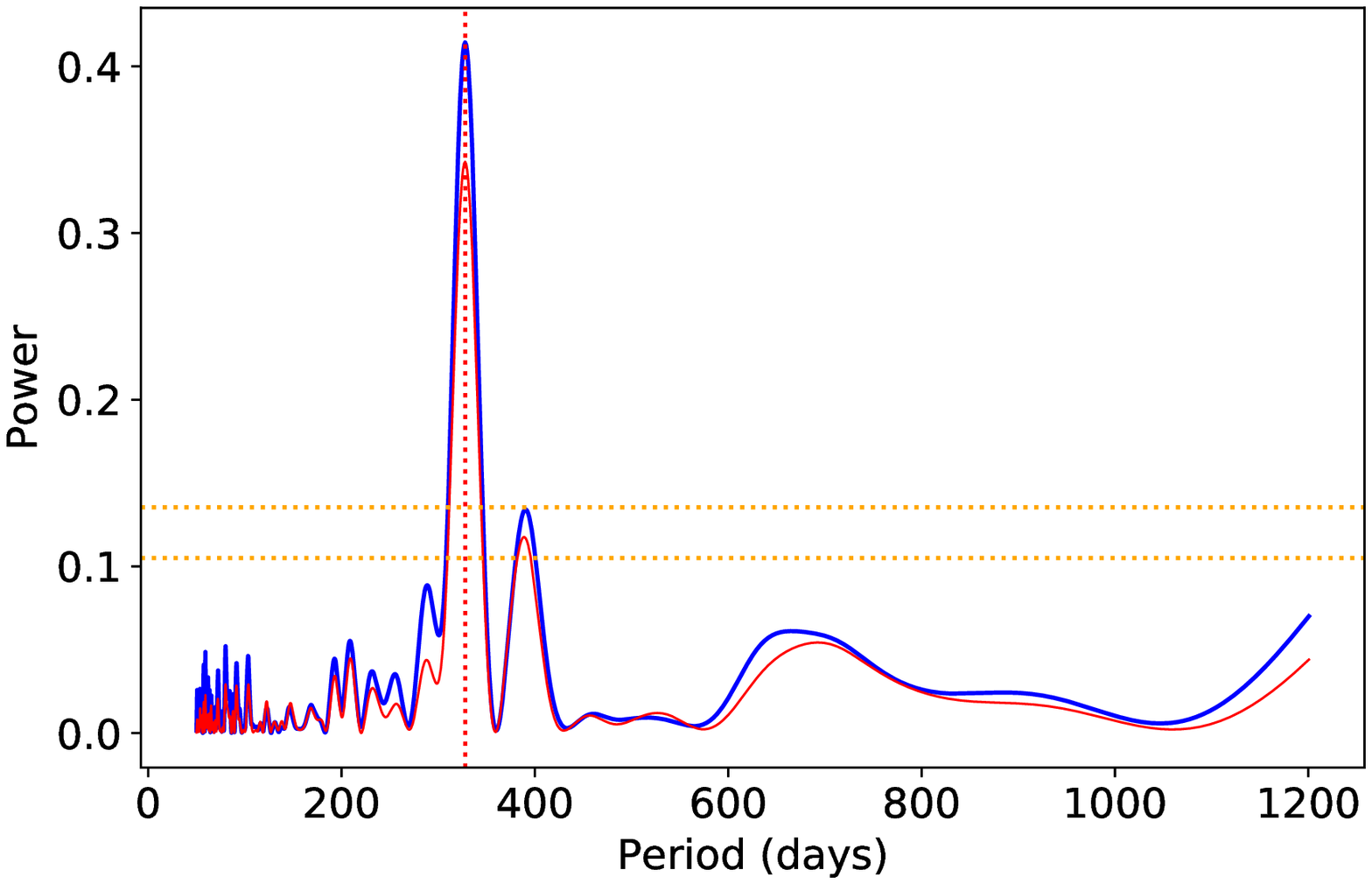}
	\centering\includegraphics[width = 7cm,height=5cm]{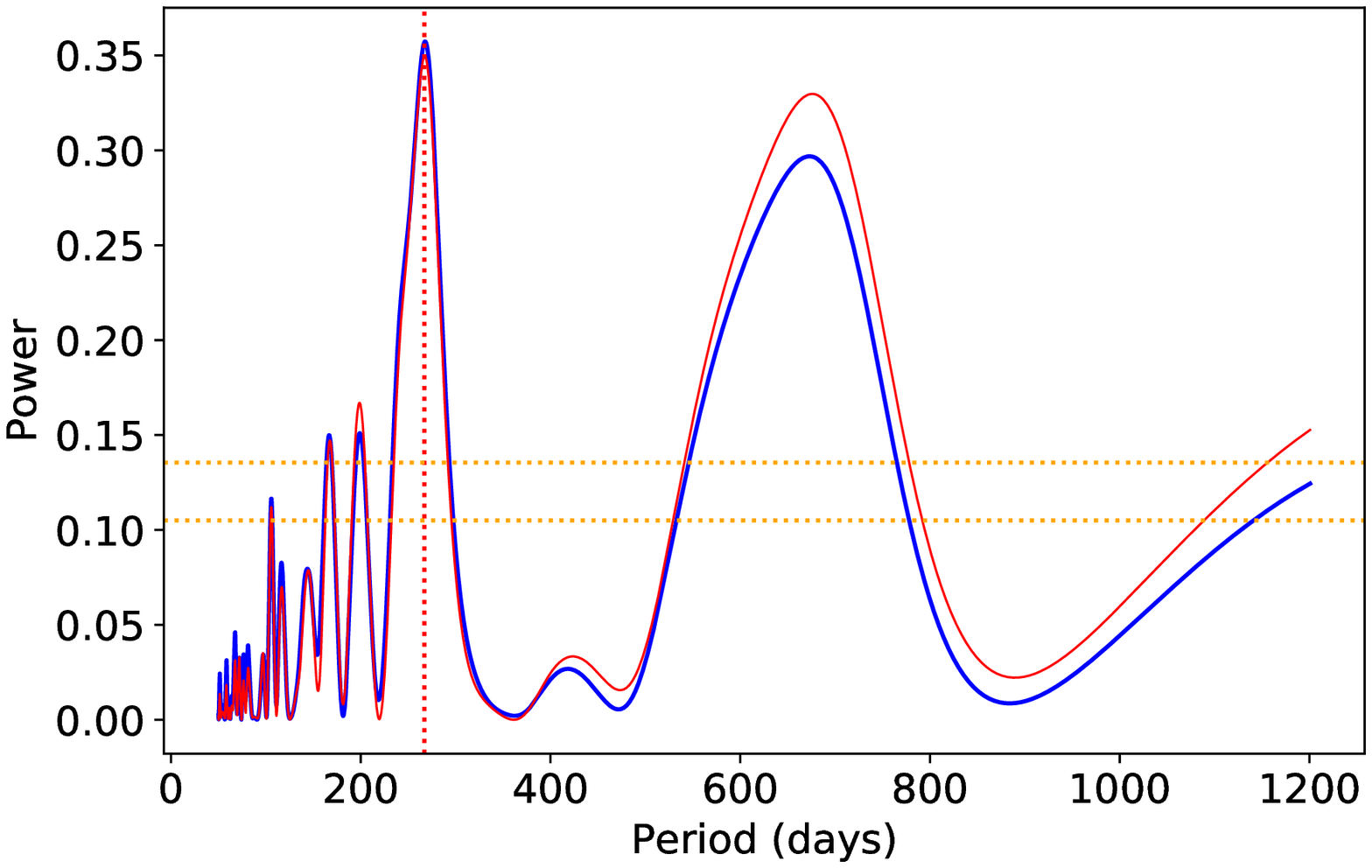}
	\caption{The powers determined through the GLSP method applied to the CSS light curve (left panel) 
		and to the ASAS-SN light curve (right panel).
		Solid blue line in each panel represents the GLSP power of CSS and ASAS-SN 
		light curves, and solid red line
		represents the GLSP power of the evenly sampled
		CSS and ASAS-SN light curves.
		The vertical red dotted line in each panel marks the 
		position of the corresponding peak of the power. The orange dotted lines represent significance level 
		at 99.99\% and 99.9999\%, respectively.
	}
	\label{fig2}
\end{figure*}

\begin{figure*}
	\centering\includegraphics[width = 7cm,height=5cm]{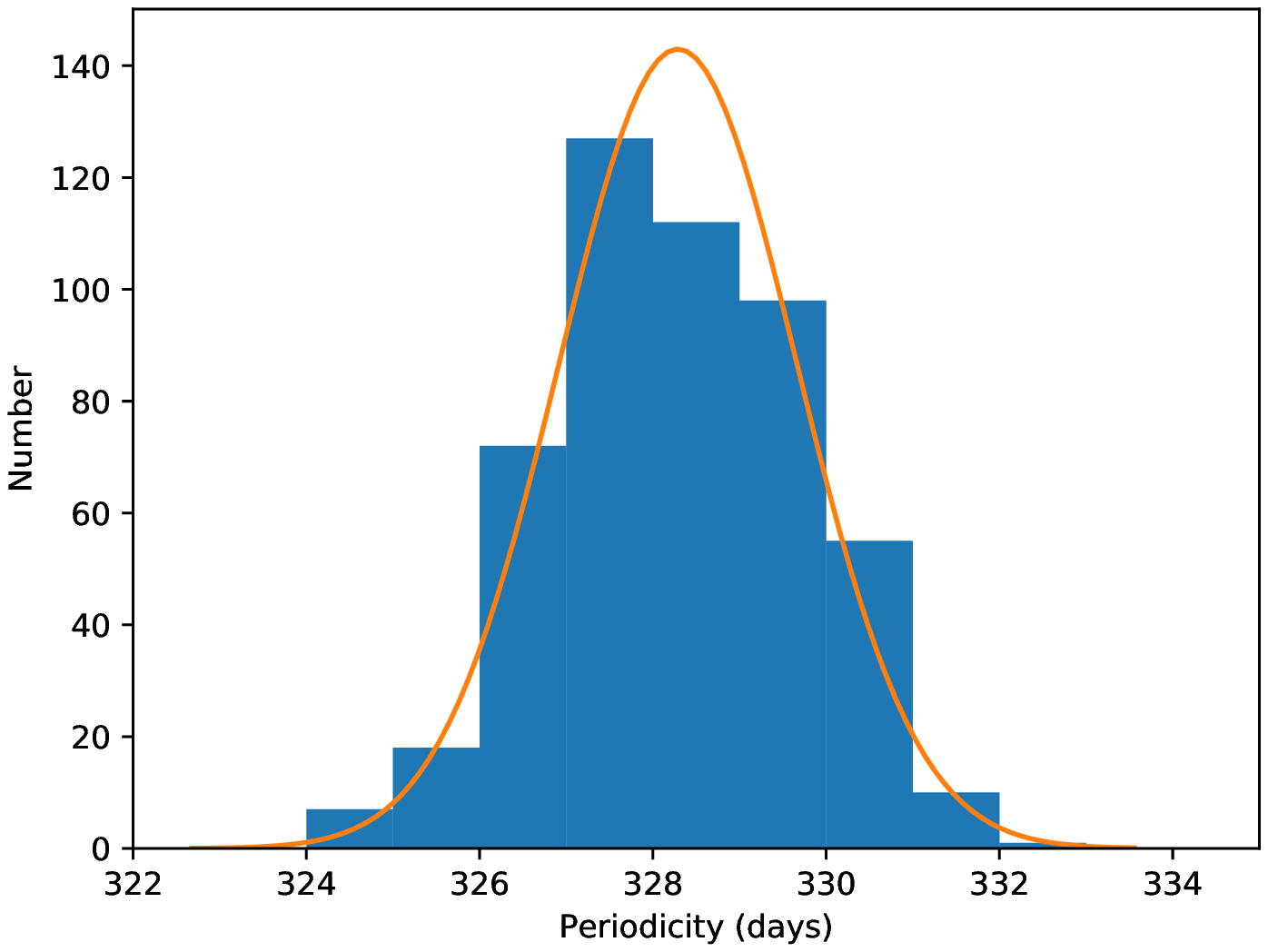}
	\centering\includegraphics[width = 7cm,height=5cm]{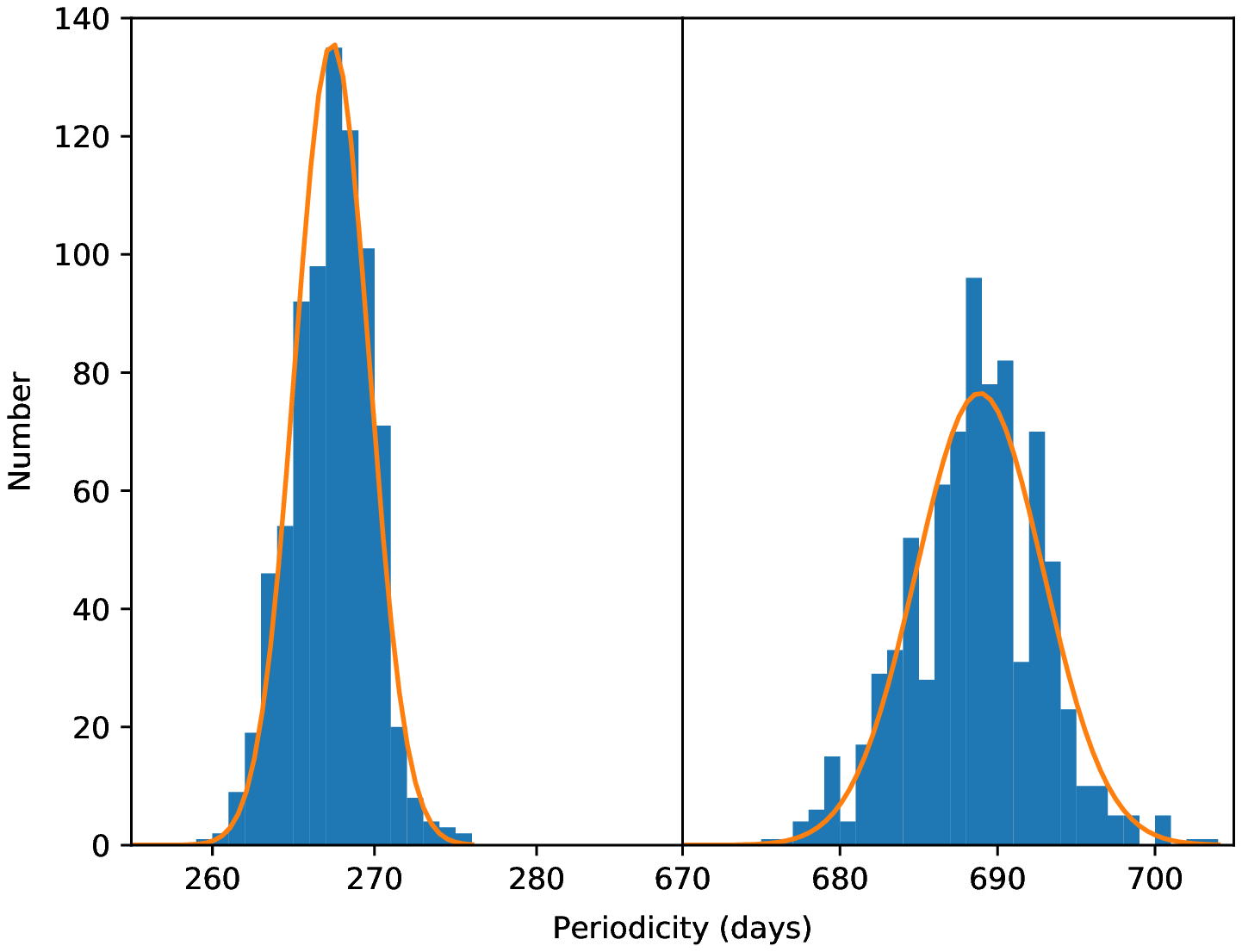}
	\caption{Left panel displays bootstrap method determined distribution of the periodicity about 
		328$\pm$4 days from CSS and right panel shows bootstrap method determined distributions of the 
		periodicity about 267$\pm$8 days and the second periodicity about 689$\pm$14 days from ASAS-SN. 
		The solid orange line in each panel marks the best Gaussian Fitting of periodicity distribution.
	} 
	\label{fig3}
\end{figure*}

Differing from the Lomb-Scargle algorithm \citep{Lo76, Sc82}, the GLSP method \citep{Ze09} 
not only considers the errors associated with the fluxes, but also uses sinusoids plus constant rather 
than sinusoidal functions as a fitting function. 
As well discussed in \citet{Ze09}, let y(t)=a~cos$\omega$t+b~sin$\omega$t+c be the fitting function and $y_i$ be the N measurements of a time series at time $t_i$ with errors $\sigma_i$. Then at given frequency $\omega$, let the squared difference between y(t) and $y_i$ minimized:
\begin{equation}
	\chi ^2=\sum\limits_{i=1}^{N}\frac{[y_i -y(t_i)]^2}{\sigma_i ^2}.	
\end{equation}	
Furthermore, the power P($\omega$) normalised to unity by $\chi_0 ^2$  ($\chi ^2$ for the weighted mean) can be written as:
\begin{equation}
	P(\omega)=\frac{\chi_0 ^2 - \chi ^2 (\omega)}{\chi_0 ^2}	
\end{equation}	
Figure \ref{fig2} shows the powers from the GLSP method. 
Through the CSS light curve, there is a clear peak around 328$\pm$4 days with significance level higher 
than 99.9999\% (false alert probability 0.000001 in GLSP). 
As discussed in \citet{Va18}, the significance is usually expressed in terms of a false alarm probability, encoding the probability of measuring a peak of a given height (or higher) conditioned on the assumption that the data consists of Gaussian noise with no periodic component.
Through the ASAS-SN light curve, there is a 
clear peak around 267$\pm$8 days with significance level higher than  99.9999\% (false alert probability 
0.000001 in GLSP). Meanwhile, there is an additional peak of 689$\pm$14 days in ASAS-SN light curve. 
And the uncertainties of the periodicities are determined by the widely applied bootstrap method 
leading to periodicity distribution shown in Figure \ref{fig3}. 
Meanwhile, based on the GLSP power properties shown in Figure 2
, quality of 300 days QPOs as discussed in \citet{gm08} can be estimated by 
$T/\delta T \sim 11.3$ ($T$ as periodicity and $\delta T$ as full width at half maximum) in CSS light curve and 5.2 in ASAS-SN light curve, indicating there is a high quality periodicity.

In addition, based on Figure \ref{fig1}, the variability amplitude in CSS light curve (standard deviation about 0.54) is about 1.6 times larger than the variability amplitude in ASAS-SN
light curve (standard deviation about 0.34). 
The definite reason of the different variability amplitudes (probably due to intrinsic variability related to central accreting process) is unknown.
But the quite different variability amplitudes have apparent effects on detecting QPOs through GLSP, probably leading to different peak values in GLSP power.
So the data of CSS and ASAS-SN are not put together.

\begin{figure*}
	\centering\includegraphics[width = 7cm,height=5cm]{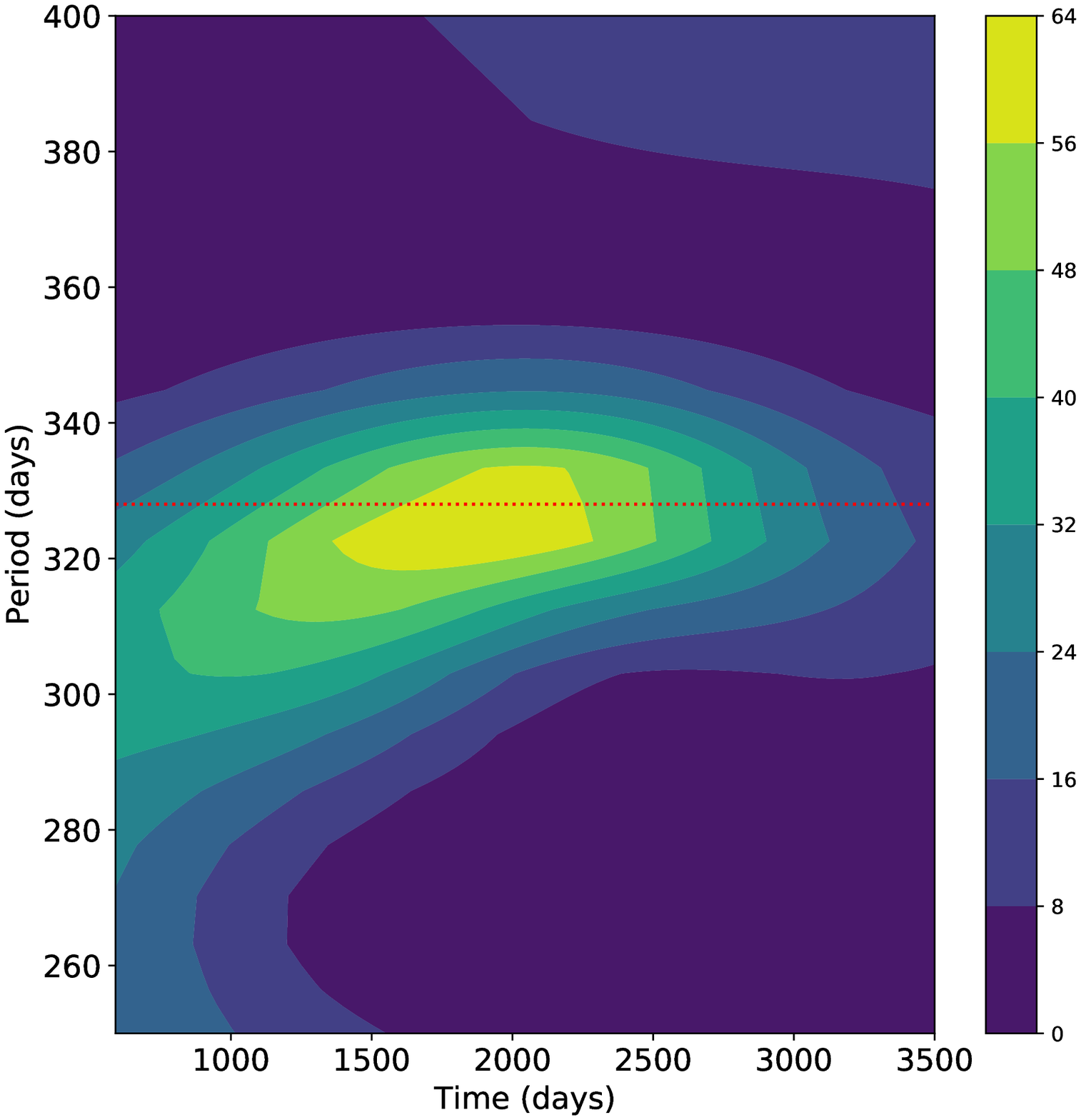}
	\centering\includegraphics[width = 7cm,height=5cm]{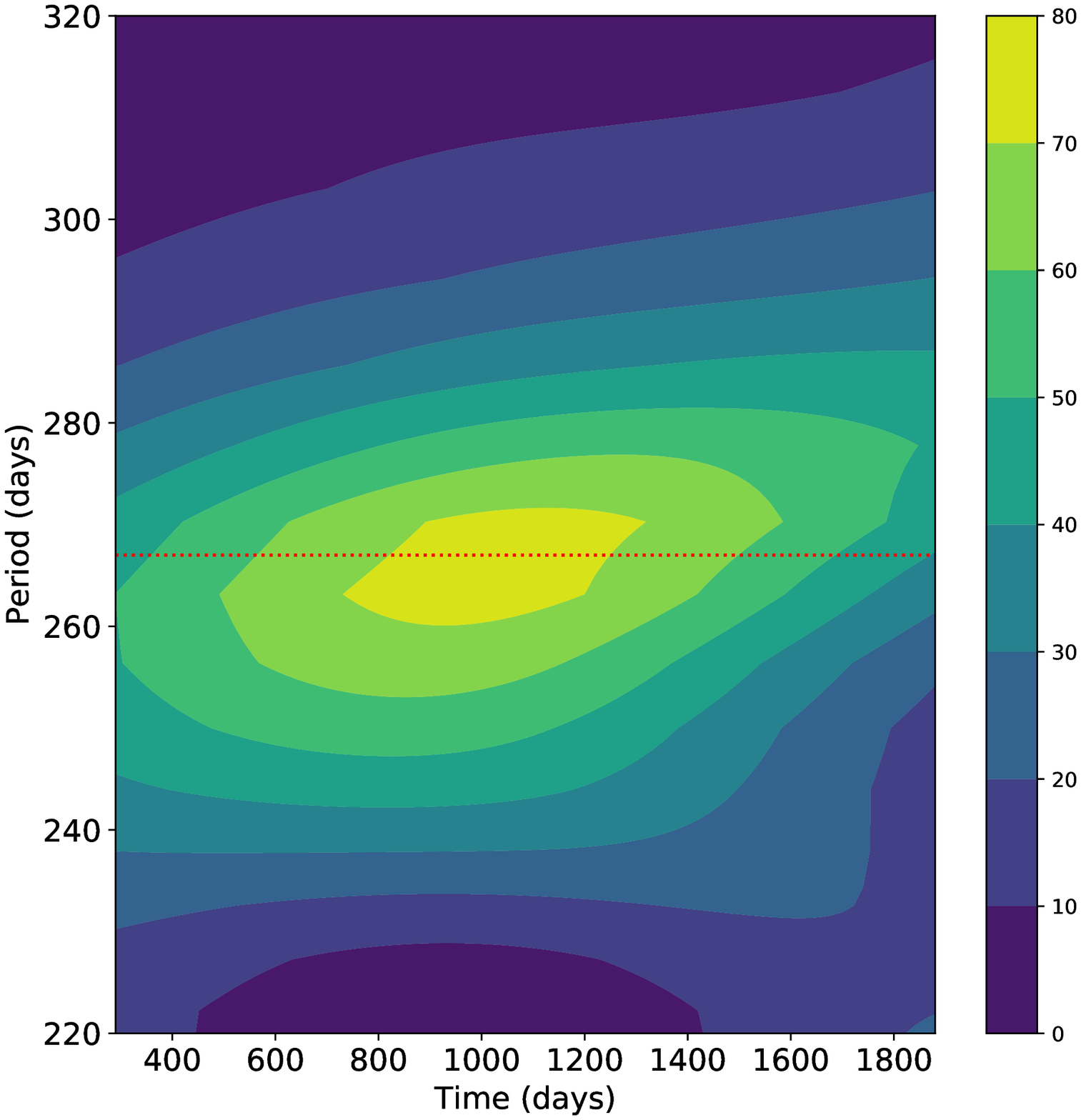}
	\caption{Power properties through WWZ applied to the optical CSS V-band light curve (left panel) and to 
		the optical ASAS-SN V-band light curve (right panel). The horizontal dotted red line in each panel marks 
		the position of the corresponding periodicity. 
	}
	\label{fig4}
\end{figure*}

WWZ firstly proposed by \citet{Fo96}  can be well applied to estimate and determine QPOs \citep{To98,An13,Li21}, especially in unevenly sampled time series,
based on three trial
	functions:${\bf 1}(t)$, $cos[\omega(t-\tau)]$ and $sin[\omega(t-\tau)]$.
	The ${\bf 1}(t)$ represents a constant function, since the function firstly described by \citet{Fo96}.
	 $w_{\alpha}=exp(-c\omega ^2({\rm t}_{\alpha}- \tau)^2)$ ($\alpha=1, 2, 3$) is the statistical weight and in which c is a tunable parameter. The WWZ power is defined with
\begin{equation}
	{\rm WWZ}=\frac{(N_{\rm eff}-3)V_{\rm y}}{2(V_{\rm x}-V_{\rm y})}.	
\end{equation}
$N_{\rm eff}$ represents the effective number density of data points, and the weighted variations of 
data x and model function y are $V_{\rm x}$ and $V_{\rm y}$, respectively.
These factors are described as:
\begin{equation}
	N_{\rm eff}=\frac{(\sum w_{\alpha})^2}{\sum w_{\alpha}^2}=\frac{[\sum e^{-c \omega^2({\rm t_{\alpha}}-\tau)^2}]^2}{\sum e^{-2c \omega^2({\rm t_{\alpha}}-\tau)^2}} ,
\end{equation}
\begin{equation}
	V_{\rm x}=\frac{\sum_{\alpha} w_{\alpha}x^2(\rm t_\alpha)}{\sum_{\lambda} w_{\lambda}}-\left[\frac{\sum_{\alpha} w_{\alpha}x(\rm t_\alpha)}{\sum_{\lambda} w_{\lambda}} \right]^2=<x|x>-<1|x>^2 ,
\end{equation}
\begin{equation}
	V_{\rm y}=\frac{\sum_{\alpha} w_{\alpha}y^2(\rm t_\alpha)}{\sum_{\lambda} w_{\lambda}}-\left[\frac{\sum_{\alpha} w_{\alpha}y(\rm t_\alpha)}{\sum_{\lambda} w_{\lambda}} \right]^2=<y|y>-<1|y>^2.
\end{equation}
In Figure \ref{fig4}, the 
powers show clear peaks at $\thicksim$328 days in the CSS light curve and at $\thicksim$ 266 days in 
the ASAS-SN light curve, respectively. The periodicities determined through the WWZ technique are well 
consistent with the results determined by the GLSP method.

\begin{figure*}
	\centering\includegraphics[width = 7cm,height=5cm]{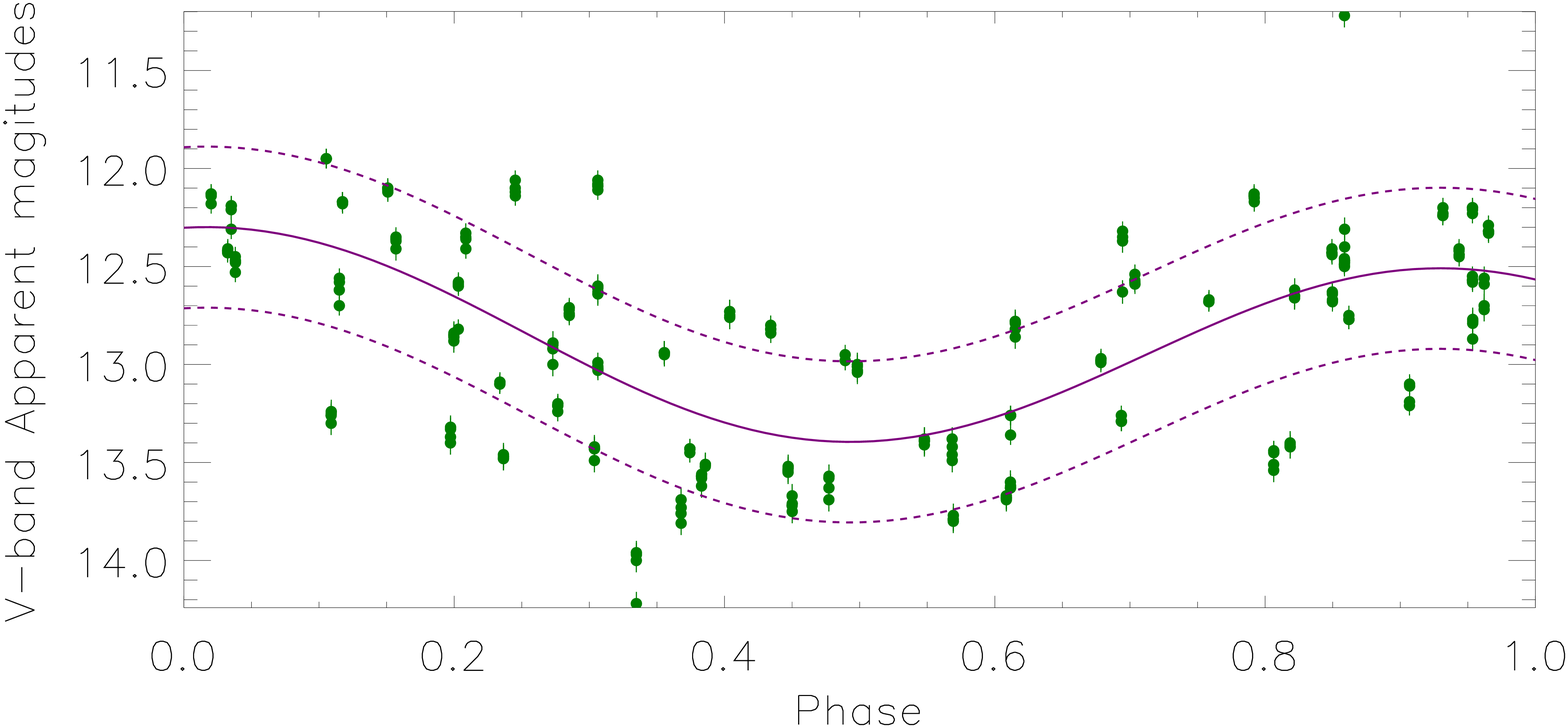}
	\centering\includegraphics[width = 7cm,height=5cm]{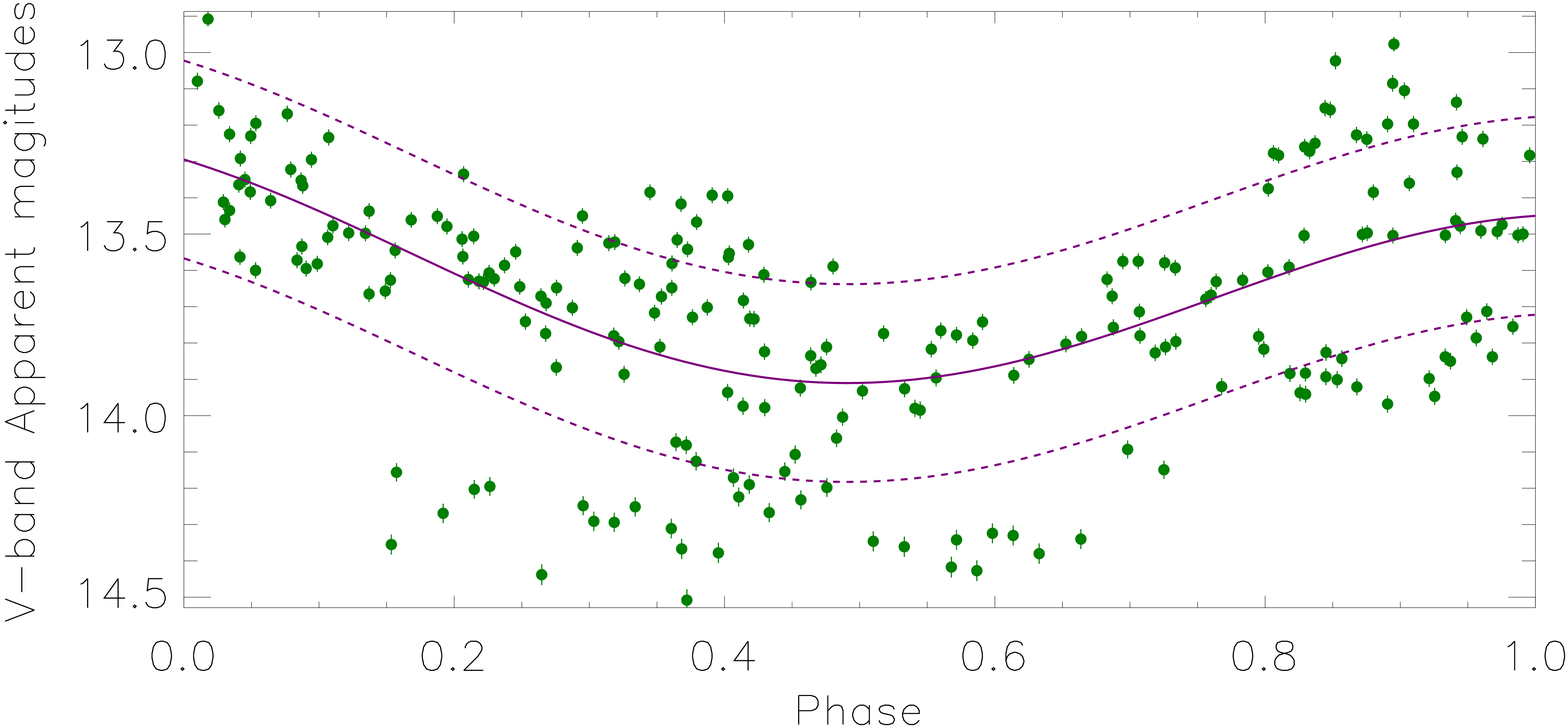}
	\caption{Epoch-folded CSS light curve with a 328 day periodicity (left panel) and epoch-folded ASAS-SN light 
		curve with periodicity of about 267 days (right panel). The solid purple lines are the best-fitting descriptions 
		by sinusoidal function, and dashed purple lines show the corresponding 1RMS scatters.} 
	\label{fig5}
\end{figure*}

\begin{figure*}
	\centering\includegraphics[width = 14cm,height=8cm]{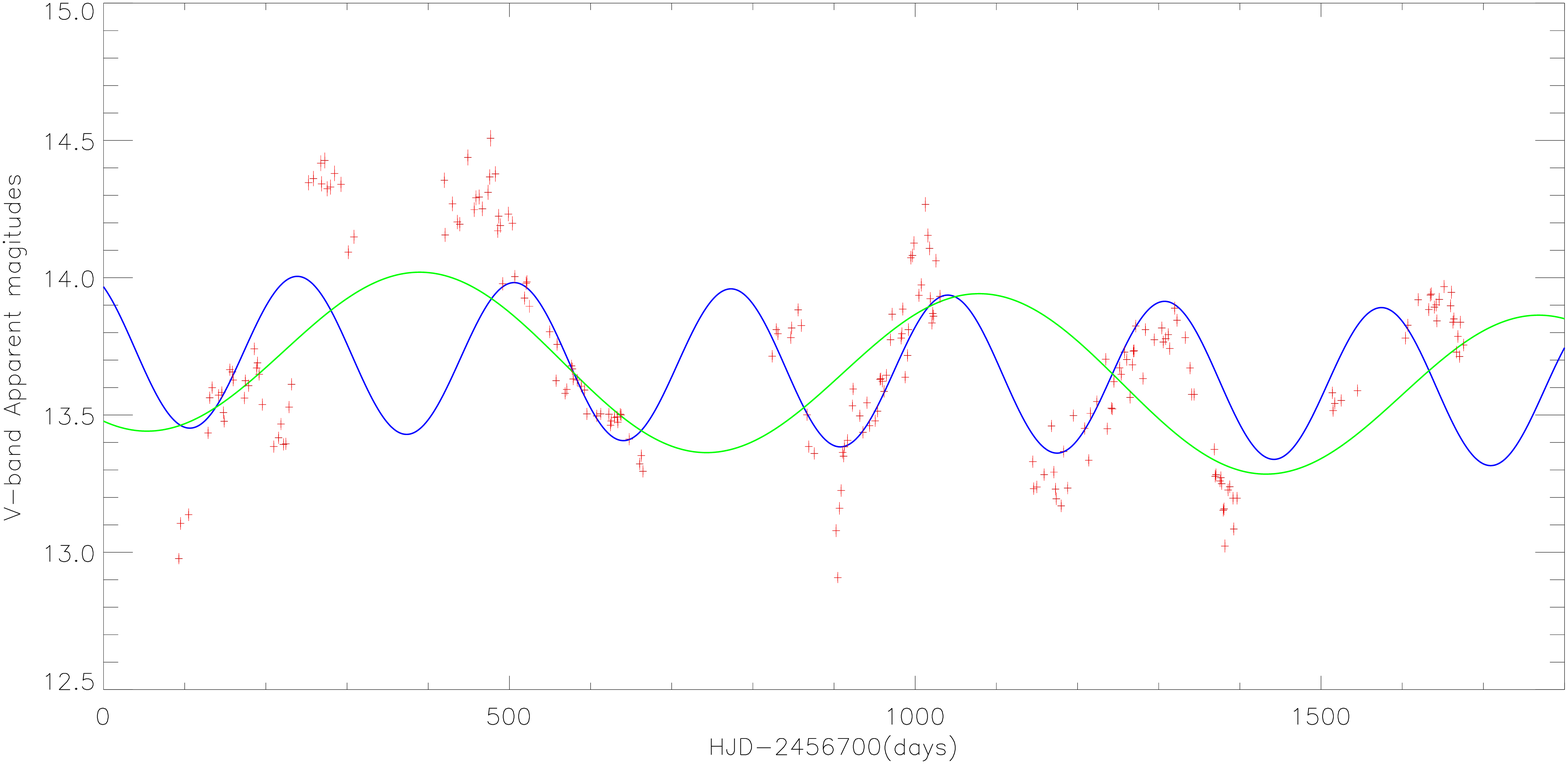}
	\caption{Light curve(red plus) of PKS~2155-304\ in V-band from ASAS-SN.
		Solid blue line shows the best fitting result with periodicity as a free model
		parameter, solid green line represents the best fitting result with fixed periodicity.
	}
	\label{fig6}
\end{figure*}

Through the CSS light curve, both GLSP and WWZ show a clear peak at about 328 days. Meanwhile, 
through the ASAS-SN light curve, besides the peak about 267 days, there is an additional second peak 
around 689 days through the GLSP method. In order to 
confirm the periodicity around 300days, the epoch-folded 
method is applied. Left panel of Figure \ref{fig5} shows the folded CSS light curve with periodicity about 
328 days and with zero point corresponding to MJD=53597.542. Right panel of Figure \ref{fig5} shows the 
folded ASAS-SN light curve with periodicity about 267 days and with phase zero point corresponding to 
HJD=2456752.777. The epoch-folded light curves can be well described by sinusoidal function shown as solid 
purple lines in Figure \ref{fig5}, to support the periodicity around 300 days. 
In order to test the second periodicity around 689 days detected in the ASAS-SN light curve 
by the GLSP method, two model functions are applied to describe the ASAS-SN
light curve. The first model function (model 1) is $a~+b\times t+c\times sin(2\pi \times t/T +\phi_0)$
with periodicity $T$ as a free model parameter, leading to the best descriptions
shown as solid blue line in Figure \ref{fig6} with $\chi_1^2/dof_1 \sim 26995.2/221$ 
(sum of squared residuals divided by degree of freedom) with determined $T\sim267 days$
totally similar as the GLSP determined first periodicity.
The second model function (model 2) is $a~+b\times t+c\times sin(2\pi \times t/689 +\phi_0)$
with periodicity $T=689 days$(the second periodicity of 689 days detected by GLSP) as a fixed model parameter, leading to the best descriptions
shown as solid green line in Figure \ref{fig6} with $\chi_2^2/dof_2 \sim 29283.1/222$.
Due to high accuracy of ASAS-SN data point, there are large $\chi^2$ value.
In addition, it is useful to determine whether periodicity of 689 days is preferred 
through the F-test technique.
Based on the different $\chi^2/dof$
values for the model 1 and Model 2, the calculated $F_p$ value is about
\begin{equation}
	F_p=\frac{\frac{\chi^2_2-\chi^2_1}{dof_2-dof_1}}{\chi^2_1/dof_1}\sim18.7.
\end{equation}
Based on $dof_2-dof_1$ and $dof_1$ as number of dofs of the F distribution numerator and denominator,
the expected value from the statistical F-test with confidence level about 0.0021\% will be near to $F_p$.
Therefore, the confidence level is higher than 99.9979\% (1-0.0021\%), higher than 3sigma, to support that periodicity of 267 days is preferred, rather than the periodicity of 
689 days.
Therefore, there are no further discussions on the periodicity of 689 days.

A simple fact is that we discover the T$\sim$328 days periodicity from data of CSS light curve and 
the T$\sim$267 days periodicity from data of ASAS-SN. The results are similar to the optical QPOs with a 
periodicity of 317 days in R-band in \citet{Zh14}, a periodicity of 315 days in VRIJHK band in \citet{Sa14} 
and a periodicity of 315 days in R-band in \citet{Sa18}, strongly supporting the expected optical QPOs with 
periodicity about 300 days in the well-known blazar PKS~2155-304.  

\begin{figure*}
	\centering\includegraphics[width = 7cm,height=5cm]{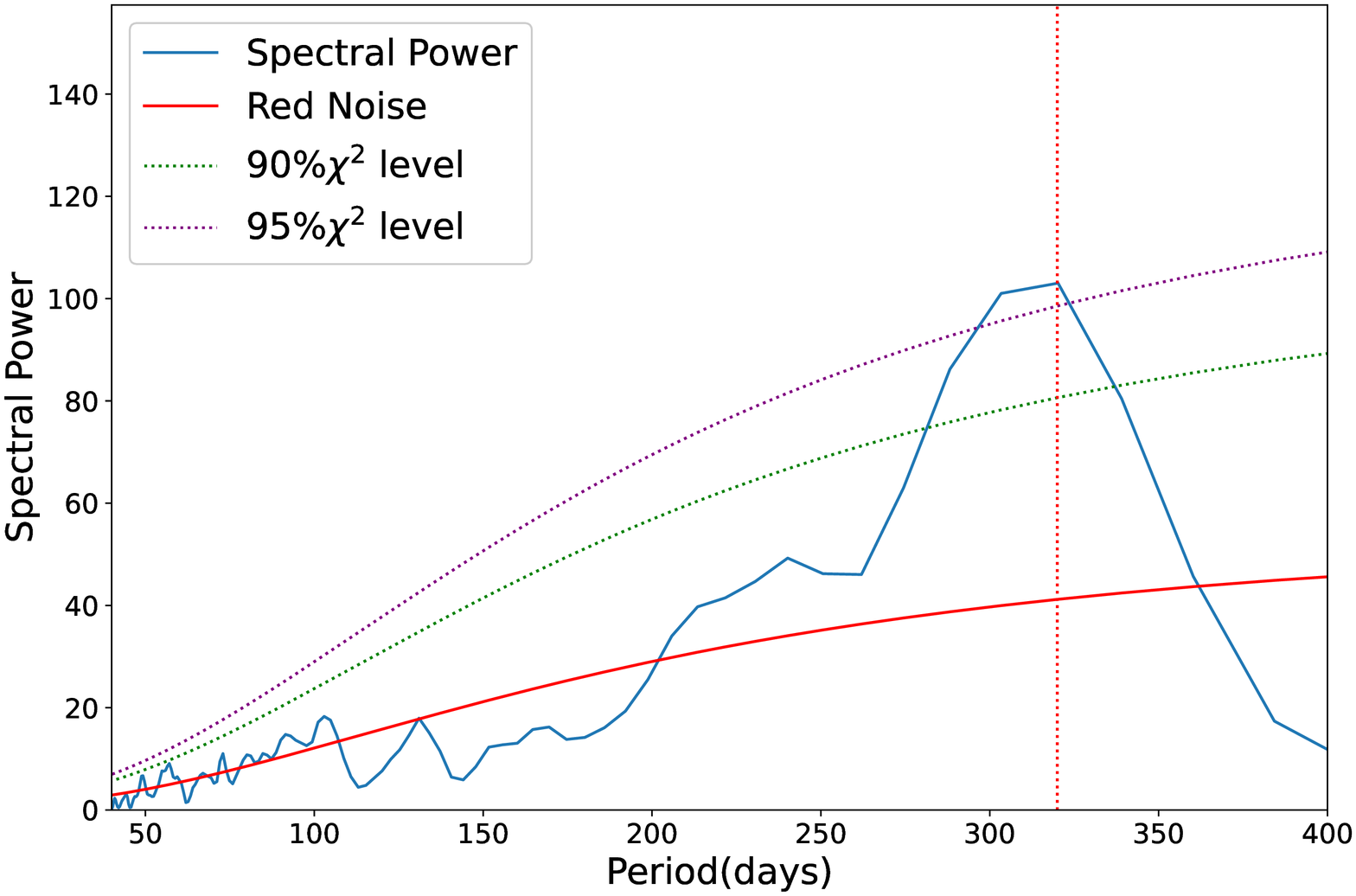}
	\centering\includegraphics[width = 7cm,height=5cm]{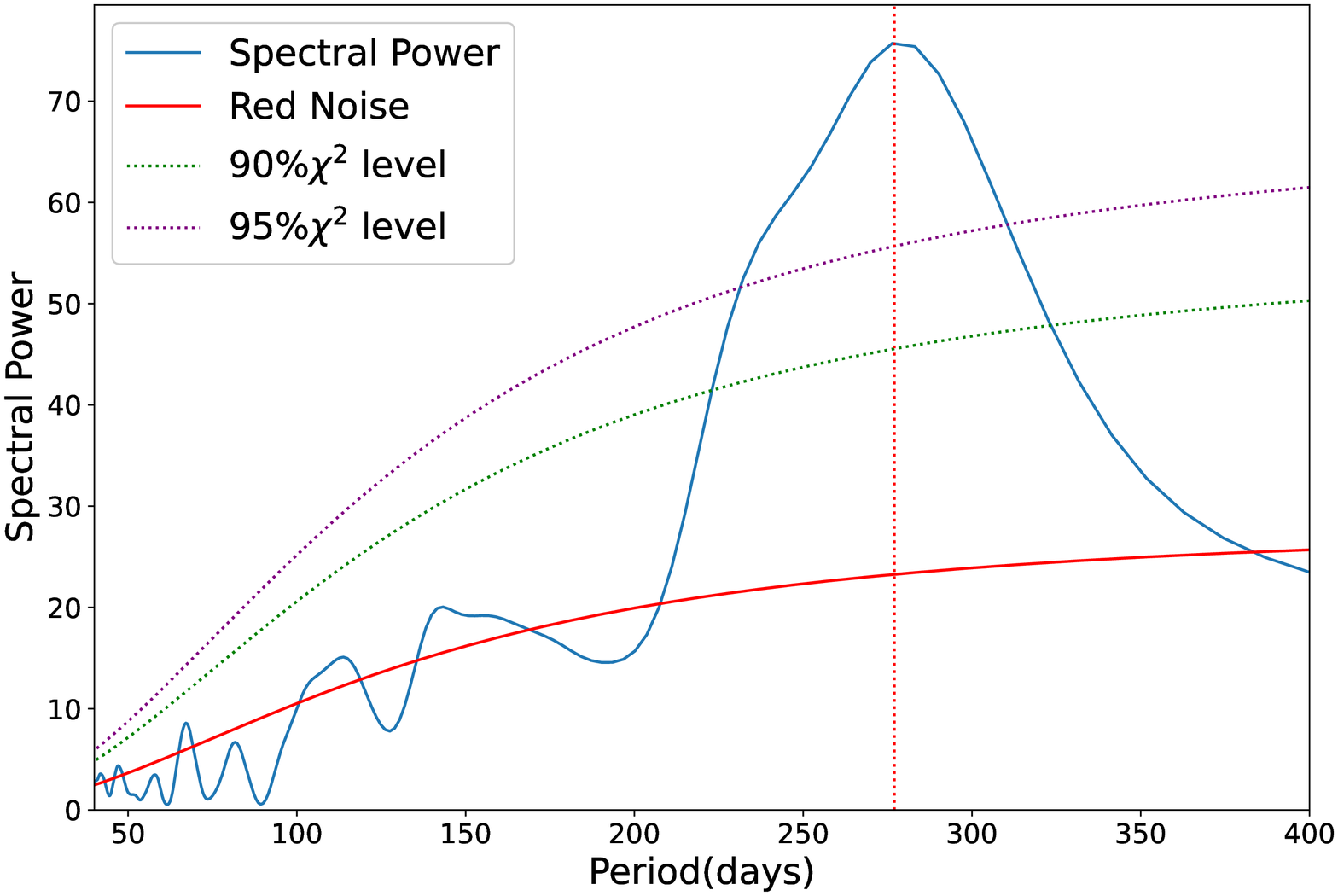}
	\caption{The power spectrum calculated by redfit method applied to CSS light curve (left panel) and ASAS-SN (right panel) light curve. }
	\label{fig7}
\end{figure*}

However, the presence of red noise at optical band can affect the detected QPOs in AGN. It not only can bury possible QPOs signal, but also may spuriously mimic few-cycle sinusoid-like periods \citep{Kr21}. \citet{Co19} has showed that some periodicities reported in AGNs appears poorly justified, but \citet{Re211} said that red noise can hardly be responsible for long-term QPOs.
In the paper, the influence of red noise is eliminated.
\Citet{Sc02} provided a computer program (redfit) which can estimate red-noise spectra from unevenly spaced data and give the confidence level. In addition, this program is based on two assumptions \citep{Xo17}: (1) the noise background can be approximated by continuous autoregressive (CAR); (2) the distribution of data points is not too clustered. Based on the redfit method, the influence of red noise in PKS 2155-304 in CSS and ASAS-SN light curves is check in Figure \ref{fig7}. It is obvious that
the determined periodicities are around 320 days and 277 days through the redfit method
applied to CSS and ASAS-SN light curves, respectively, similar as the results by GLSP, WWZ and epoch-folded method. 
What is more, the confidence levels of the periodicity in CSS and ASAS-SN light curves are both much higher than 95\%.

\begin{figure*}
	\centering\includegraphics[width = 7cm,height=5cm]{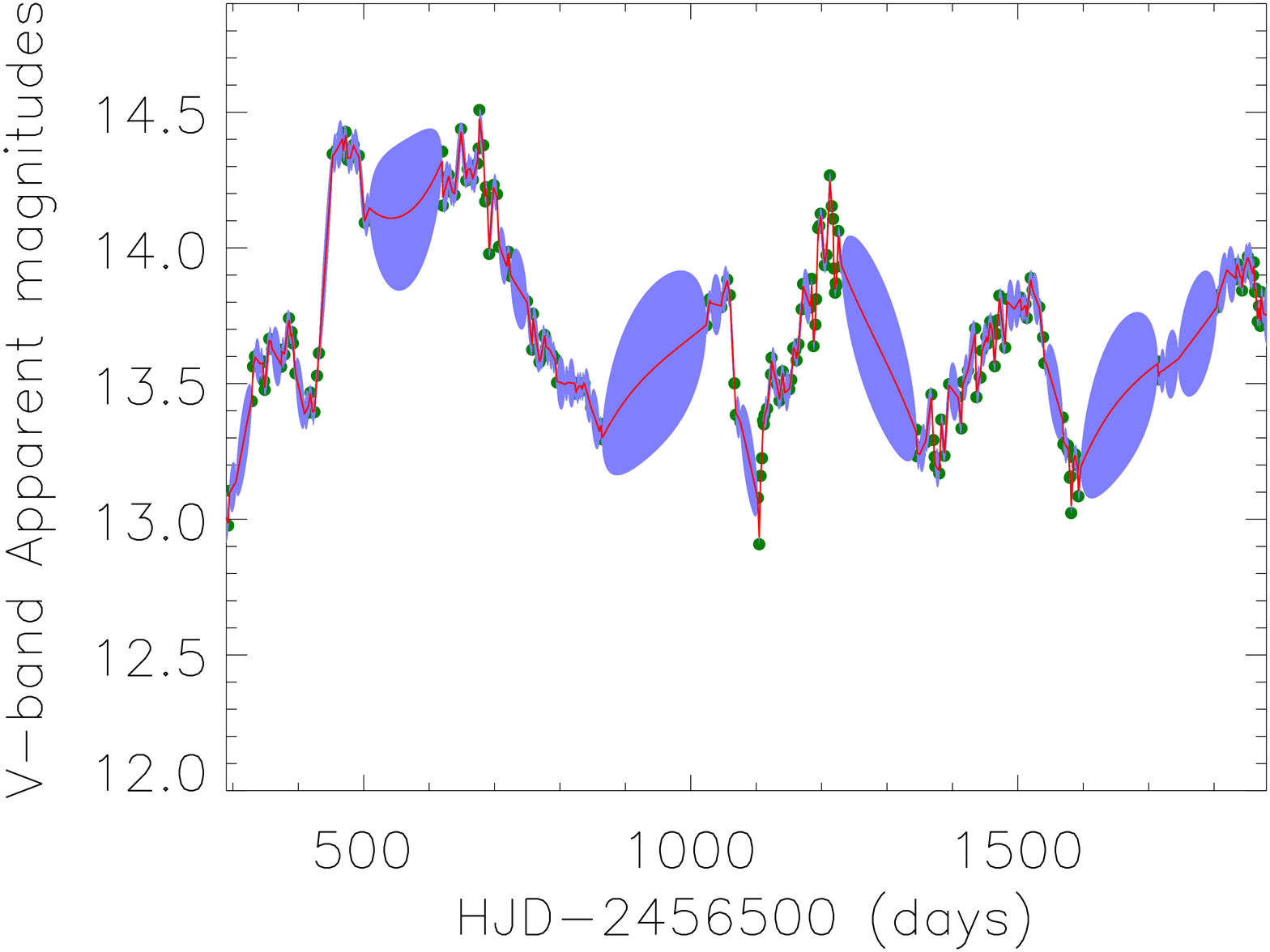}
	\centering\includegraphics[width = 7cm,height=5cm]{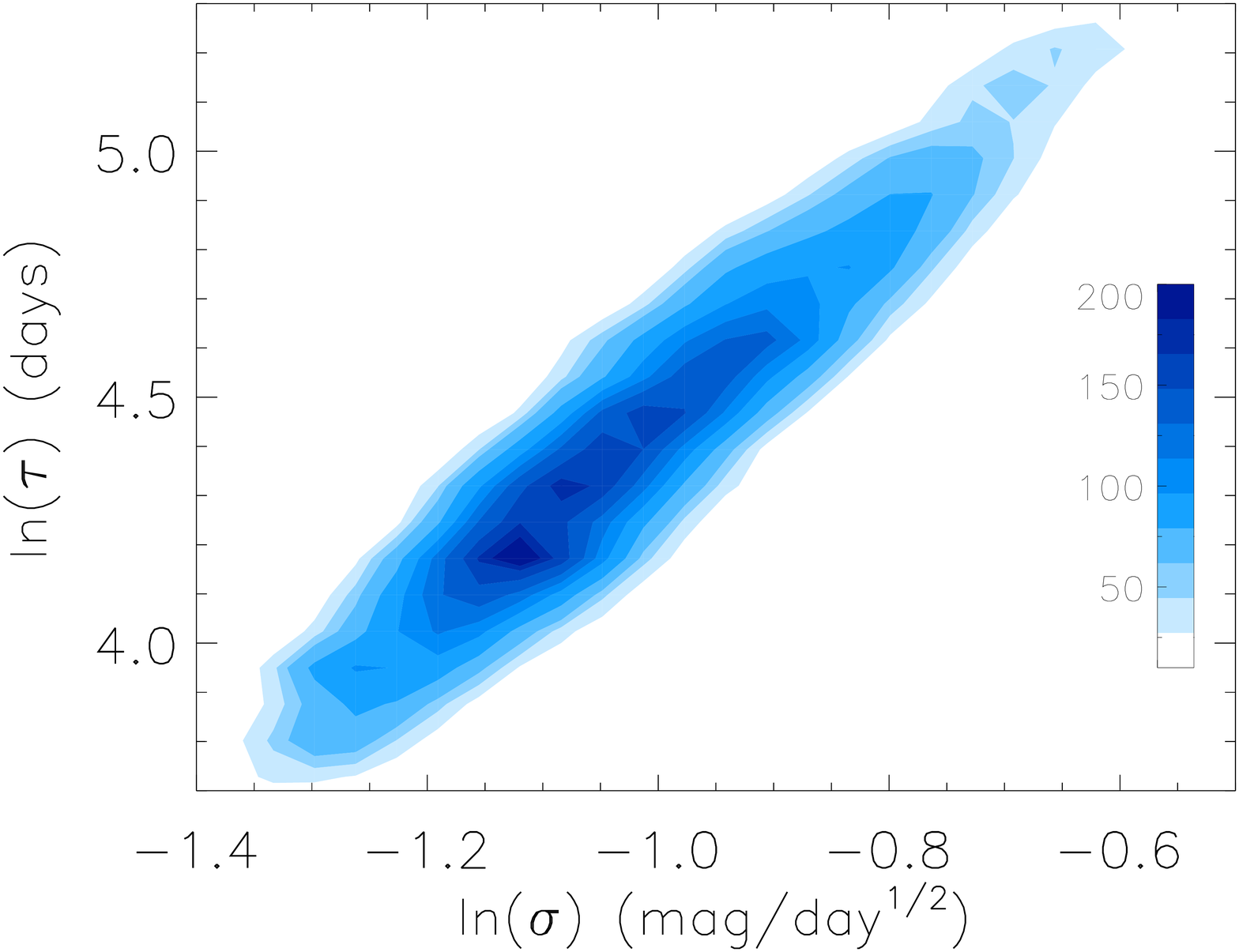}
	\caption{The DRW-determined best descriptions to the light curve of PKS 2155-304 (left panel) and the posterior distributions of $\tau$ and $\sigma$ (right panel).
Solid red line represents the best descriptions and area filled with light blue shows the corresponding 1$\sigma$ confidence bands and solid green dot is the data of PKS 2155-304 from ASAS-SN in the left panel. 
}
	\label{fig8}
\end{figure*}

Moreover, the damped random walk (DRW) process and/or continuous autoregressive (CAR) are used to examine the probability of detecting fake QPOs.
The CAR process is discussed in \citet{Ke09}:
\begin{equation}
	dX(t)=- \frac{1}{\tau}X(t)+ \sigma \sqrt{dt} \epsilon (t)+bdt, \tau , \sigma , t >0,
\end{equation}
where $\tau$ and $\sigma$ are intrinsic characteristic variability amplitude and timescale, respectively.
$\epsilon(t)$ is a white noise process, and X(t) is the AGN light curve and $\tau \sigma ^2/2$ is the variance.
According to the public code JAVELIN (Just Another Vehicle for Estimating Lags In Nuclei) \citep{Ko10,Zu13}, the $\tau$ of light curve in CSS and ASAS-SN is 132$\pm$55 days and 83$\pm$36 days, respectively. 
Here, due to higher quality of ASAS-SN light curve, only the best descriptions to ASAS-SN light curve is shown in the left panel of Figure \ref{fig8}.
Moreover, the effects of insufficient data sampling can be simply discussed.
Based on the best descriptions, an evenly sampled light curve 
(time step about 0.8 days) can be determined and shown as solid red line in left panel 
of Figure \ref{fig8} .
And then, solid red line in the right panel of Figure \ref{fig2} shows the GLSP power properties determined from the evenly sampled ASAS-SN light curve, similar as the GLSP power properties from observed ASAS-SN light curve.
Similar results can be found in the left panel of Figure \ref{fig2} to the observed CSS light curve and the evenly sampled CSS light curve (time step about 0.2 days).
So that, the insufficient data sampling has few effects on our final results.

\begin{figure*}
	\centering\includegraphics[width = 7cm,height=5cm]{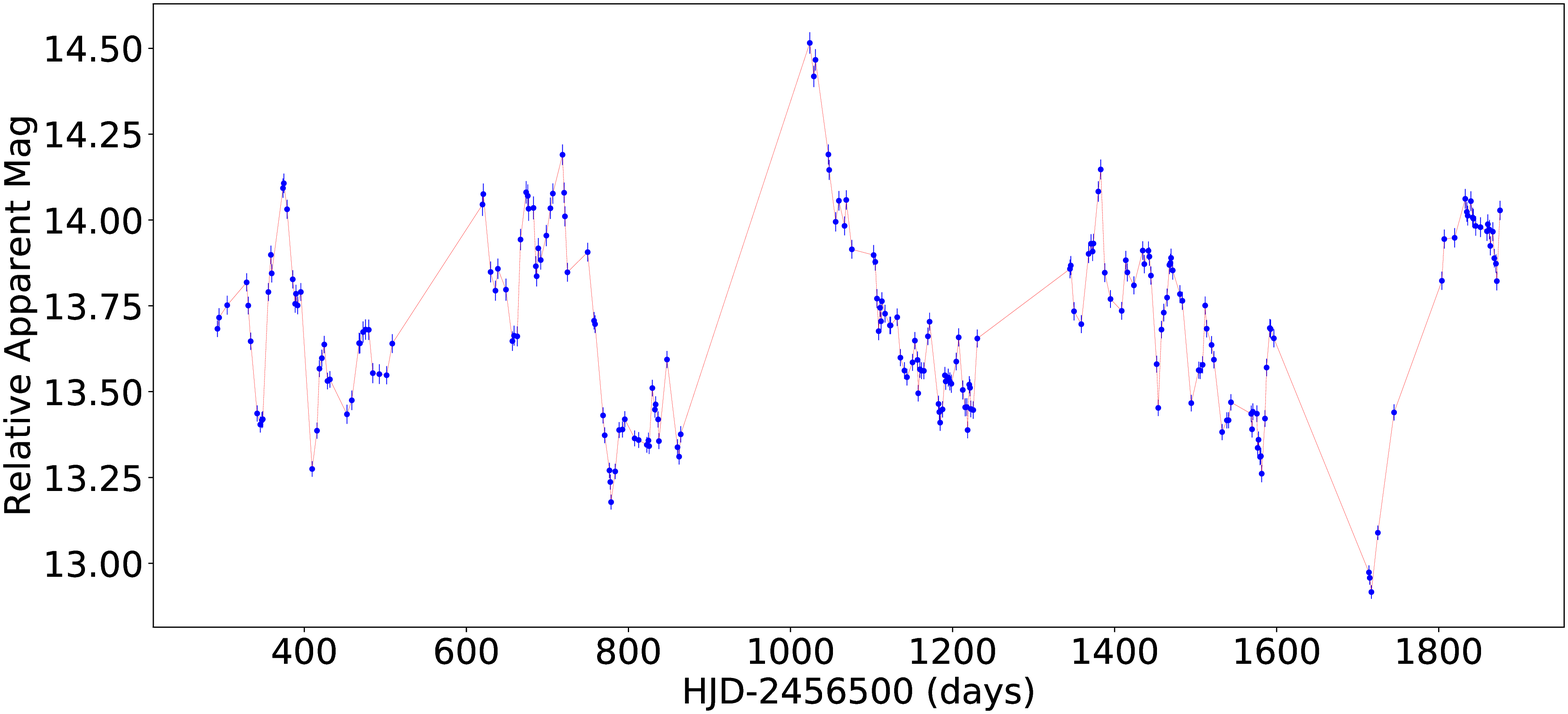}
	\centering\includegraphics[width = 7cm,height=5cm]{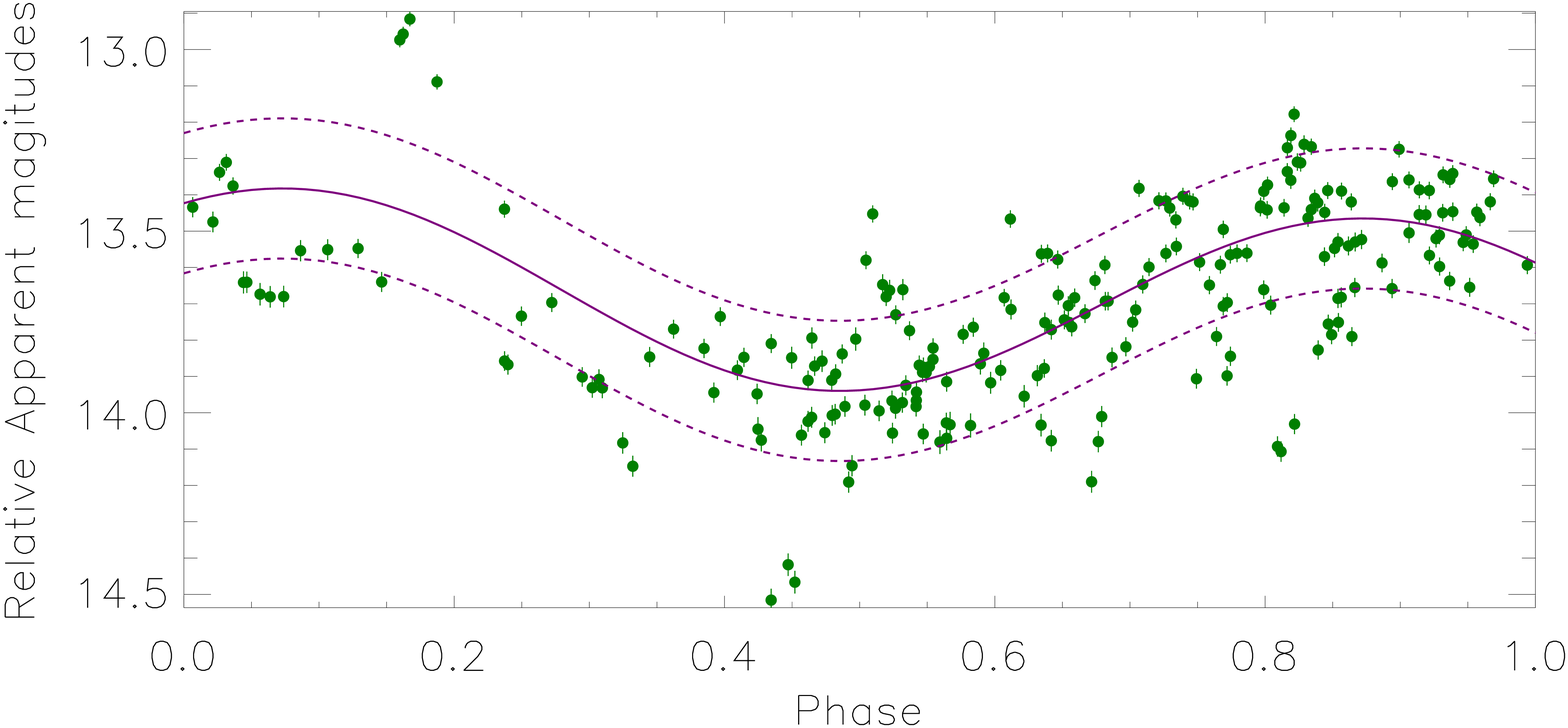}
	\caption{An example of light curve with fake QPOs by CAR process (left panel) and corresponding 
 epoch-folded light curve with periodicity about 400 days (right panel).
The solid purple line represents the best-fitting descriptions by sinusoidal function, and dashed purple lines show the corresponding 1RMS scatters in the right panel.
} 
	\label{fig9}
\end{figure*}

And the posterior distributions of $\tau$ and $\sigma$ determined by Markov Chain Monte Carlo (MCMC) \citep{Fo13} technique are presented in the 
right panel of Figure \ref{fig8}.
Then it is interesting to determine whether fake QPOs can be detected in CAR process created light curves.
Based on the equation (8), the 1000 artificial light curves are created by the following four steps.
	First, the CAR process parameters are accepted as $\tau=132 days$ (83 days), $\tau \sigma ^2/2=0.29$ (0.29 as the variance of the CSS light curve)($\tau \sigma ^2/2=0.11$ with 0.11 as the variance of the ASAS-SN light curve) and $bdt=12.87$ (12.87 as the mean value of the CSS light curve)($bdt=13.69$ with 13.69 as the mean value of the ASAS-SN light curve).
	Second, time information $t$ is the same as that of the CSS (ASAS-SN) light curve.
	Third, the white noise $\epsilon(t)$ here is randomly created
	with $\sc{randomn}$ function in IDL.
	Fourth, after 1000 loops, 1000 artificial light curves can be created by CAR process.
Then fake QPOs are detected among the artificial light curves based on the 
following three criteria.
First, the light curve can show well QPOs with GLSP determined peak value higher than 0.3 (
peak values in Figure \ref{fig2} higher than 0.35) ; 
Second,  the periodicity of the light curve is between 100 and 500 days;
Third, the light curve can be well fit with epoch-folded method.
Among the CAR process simulated 1000 light curves with $\tau$ and $\sigma$ determined
from the CSS light curve (ASAS-SN), there are four light curves (eight light curves)
with determined QPOs based on the three criteria above.
The probability of detecting fake QPOs is 0.4\% in CSS and 0.8\% in ASAS-SN light curve.
In addition, a CAR process created light curve with fake QPOs is shown 
as an example in the left panel of Figure \ref{fig9}, and corresponding 
epoch-folded light curve shown in the right panel of Figure \ref{fig9} with best fitting results
by sinusoidal function.

\section{Discussions}

There are several theoretical models which can be applied to explain the optical QPOs in PKS~2155-304. 
In this paper, the relativistic Frame-dragging effect, the binary black hole model and the jet precession 
model are mainly discussed as follows.

Relativistic Frame-dragging effect has been primordially applied to describe the QPOs in galactic 
X-ray binaries with central spinning black holes \citep{St98, Cu98, In19}. It is a general relativistic 
effect related to central accretion discs \citep{Fu09, In19, Zxg211}, and  it can be also applied in blazars 
\citep{Ma08, Liu21}. 
As more recent discussions in \citet{Bh21}, "the
rapidly spinning supermassive black hole can warp
spacetime and give rise to the precession of the disk
owing to the Lense–Thirring precession."
The Lense-Thirring precession frequency \citep{Cu98} is given by
\begin{equation}
	{\nu}_{\rm LT}=6.45{\times}10^4{\lvert}a_*{\rvert}\left(\frac{M}{M_\odot} \right)^{-1}
	\left(\frac{r}{r_g} \right)^{-3} \,{\rm Hz}	.	
\end{equation}
In this equation, $a_*$ is a angular momentum parameter. $M$ represents the mass of black hole, 
and $r$ is the radii of emission regions and $r_g = GM/c^2$ is the Schwarzschild radii.
Moreover, \citet{gm08} reported that RE~J1034+396 has a x-ray periodicity of  1h. 
Meanwhile, the central BH mass of RE~J1034+396 reported in \citet{cz16,jd20} is about $5\times 10^{6}{\rm M_\odot}$. 
RE J1034+396 is chosen as a standard case with the Lense-Thirring precession determined X-ray QPOs from the direct vicinity (about 10 Schwarzschild radii) of the black hole as discussed in \citet{gm08}. 
Meanwhile, if simply accepted that the optical QPOs of PKS2155-304 are also related to the Lense-Thirring precession,
	the central BH mass of PKS 2155-304 can be estimated, assumed
	optical emission regions are about 100 $\sim$ 200 
	Schwarzschild radii.

The central BH mass of PKS~2155-304 with optical periodicity about 
300 days would be $4.5{\times}10^6{\rm M_\odot}$ $\sim$ $3.6{\times}10^7{\rm M_\odot}$ with 
\begin{equation}
	M_{\rm BH}~\sim~\frac{\rm 300\,days}{1 \, \rm hour}\times5\times10^6{\rm M_\odot}(\frac{r_{\rm opt}}{10})^{-3}.
\end{equation}
And the mean mass can be roughly estimated as $1.5\times10^7{\rm M_\odot}$, which is similar as the BH mass 
in PKS~2155-304 reported in \citet{La09} through X-ray QPOs properties.  However, the estimated BH mass is 
quite smaller than $10^9{\rm M_\odot}$ determined by host galaxy absolute magnitude of PKS~2155-304 
as discussed in \citet{Fa91,Ko98,Sa14}, which is about two magnitudes higher than our estimated BH mass, probably due to the following 
two reasons.
On the one hand, according to \citet{Ko13}, the relation between
black hole mass and magnitude has large intrinsic scatter about 0.31.
On the other hand,
	the beaming effects would affect the host galaxy absolute magnitude, considering that a simple PSF function should be not
	efficient enough to totally describe central emissions including beaming effects.
If there were independent methods applied to determine the central 
BH mass of PKS~2155-304, it would provide further clues to support whether the relativistic Frame-dragging 
effect is preferred in the PKS~2155-304.


Besides the relativistic Frame-dragging effect, the binary black hole (BBH) model can be applied 
to explain the optical QPOs in PKS~2155-304, as the reported long-term optical QPOs in \citet{Br20, Ya21, Liao21, ON21}.
Under the framework of BBH model, the space separation $A_{\rm BBH}$ of the central binary black hole should be
\begin{equation}\label{eq2}
	A_{\rm BBH}=0.432{\times}M_{8}{\times}\left(\frac{P_{\rm BBH}/{\rm year}}{2652M_{8}} \right)^{\frac{2}{3}}	,
\end{equation}
where $M_{8}$ is the central total BH mass in unit of $10^8{\rm M_\odot}$ and $P_{\rm BBH}$ is the QPOs periodicity. If  
$M_{\rm BH}\sim 10^9{\rm M_\odot}$, determined by the absolute magnitude of host galaxy of PKS~2155-304, is 
accepted as the total BH mass, the expected space separation is about 
\begin{equation}
	A_{\rm BBH}~\sim~0.004\,{\rm pc}~\sim~4.8\,\,{\rm light-days}.
\end{equation} 
So small space separation indicates that it is hard to spatially resolve central binary black hole 
in central regions of PKS~2155-304\ in the near future.
However, so small space separation in BBH model is also reported in literature, such as \citet{Li211} used BBH model to explain QPOs with periodicity of 850 days in OT 081, with 
estimated space separation about 0.0076 pc.

Jet procession \citep{Ca13, Ma18, Re21} is also a possible explanation for the QPOs shown in the 
long-term light curves.  It assumes helical structure in a relativistic jet \citep{Bh18,Zh21}. 
Brightness of source varies with the viewing Angle \citep{Zh14,Li17}.
This model has been successfully used to other AGNs. \citet{Bh16} reported that OJ~287 has a 
periodicity of 400 days. \citet{Ca17} reported that PG~1553+113 
has a periodicity of 2.24 yr at 15 GHz from data of the  MOJAVE/2 cm Survey Data 
Archive (during 2009-2016). \citet{Sa21} reported that 3C 454.3 shows a periodicity of 
47 days in $\gamma$-ray and optical band. \citet{Tr21} reported that AO~0235+164 has a 
periodicity of 965 days with the radio band data from  the University of Michigan Radio Astronomical 
Observatory during 1980-2012. \citet{ZPF21} reported that J0849+5108 shows a periodicity of 176 days 
at the 15GHz from observations by the Owens Valley Radio Observatory. 
It is interesting to consider jet procession model in PKS 2155-304.

There are the 0.7 day periodicity 
in ultraviolet band \citep{Ur93} and 4.6 h periodicity in X-ray
band \citep{La09}, which are quite different from ~300 days optical periodicity of PKS 2155-304 shown
in our paper. 
Considering the differences of the observation time and the radiation area between optical band and other bands, 
it is possible to detect different QPOs in different bands. 
Therefore, more efforts should be necessary to check the jet procession model to 
explain the QPOs in PKS 2155-304.

\section{Summaries and Conclusions}

The main summaries and conclusions are as follows.
\begin{itemize}
	\item Through the CSS and ASAS-SN light curves, optical QPOs with periodicity about 300 days are reported in 
	PKS~2155-304, which are well consistent with previously reported optical QPOs in the literature 
	through light curves in different optical bands and from different projects, providing strong 
	evidence to support the optical QPOs with periodicity about 300days in PKS~2155-304.
	\item The QPOs in CSS light curve have about 8.8 cycles and in ASAS-SN light curves have about 6 cycles.	
	\item If the relativistic Frame-dragging effect is applied to explain the optical QPOs in PKS~2155-304, 
	the estimated central BH mass should be about $10^7{\rm M_\odot}$, which is quite smaller than that 
	estimated by absolute magnitude of host galaxy of PKS~2155-304.
	\item If the BBH model is applied to explain the optical QPOs, the total BH mass estimated by absolute 
	magnitude of host galaxy would lead the central space separation about 0.004 pc between the central two 
	black holes.  
	\item The optical 300 days QPOs are different from that in ultraviolet and X-ray bands found in previous references. If the jet precession model is applied, different bands may have similar QPOs. However, it is possible to detect different QPOs in different bands due to different observation time and radiation area.
\end{itemize}

\section*{Acknowledgements}
We gratefully acknowledge our referee for reading our paper again carefully and patiently, 
and re-giving us constructive comments and suggestions to greatly improve the paper.
This work is supported by the National Natural Science Foundation of China (Nos. 11873032, 
12173020). This paper has made use of the data from the CSS and ASAS-SN projects developed rapidly 
moving Near Earth Objects. The CSS web site is (\url{http://nesssi.cacr.caltech.edu/DataRelease/}). 
The paper has made use of the data from the ASAS-SN (\url{https://asas-sn.osu.edu/}). 
The web site of JAVELIN code is(\url{https://github.com/nye17/javelin/}).

\end{document}